\documentclass[reprint,amsmath,amssymb,aps,prx,longbibliography,floatfix]{revtex4-2}

\usepackage[T1]{fontenc}
\usepackage{graphicx}
\graphicspath{{Figures/}}
\usepackage{bm}
\usepackage{booktabs}
\usepackage{microtype}
\usepackage[ruled,vlined,linesnumbered]{algorithm2e}
\usepackage{xcolor}
\usepackage{hyperref}
\hypersetup{colorlinks=true,citecolor=blue,linkcolor=blue,urlcolor=blue}
\definecolor{revisionpurple}{rgb}{0.38,0.18,0.63}
\newcommand{\blue}[1]{\textcolor{black}{#1}}

\begin{document}

\title{Direct Adaptive Certification of High-Dimensional Entanglement with Bell Tests}

\author{Xu Kang Tan}
\thanks{These authors contributed equally to this work.}
\author{Jesvita Menezes}
\thanks{These authors contributed equally to this work.}
\author{Sanjan D. Murthy}
\author{Adetunmise C. Dada}
\email{adetunmise.dada@glasgow.ac.uk}
\affiliation{School of Physics and Astronomy, University of Glasgow, Glasgow G12 8QQ, United Kingdom}
\date{\today}

\begin{abstract}
Entangled photons play a crucial role in many quantum applications, and the ability to determine whether a state is entangled, and to characterise the nature of that entanglement, is vital if it is to be harnessed effectively. High-dimensional entangled states offer richer possibilities, but their additional measurement degrees of freedom make them increasingly demanding to characterise. However, adaptive Bell-test methods based on complex simultaneous perturbation stochastic approximation (CSPSA) have so far focused mainly on qubits.
Here we numerically investigate a Bell-inequality-violation-based method for detecting entanglement in unknown quantum states. We extend CSPSA to high-dimensional Bell testing by using the Collins--Gisin--Linden--Massar--Popescu (CGLMP) inequality for bipartite qudits. The resulting protocol can detect Bell-nonlocal correlations in unknown entangled states, whether pure or mixed, without first reconstructing their density matrix. Using 100 optimisation iterations in each of 100 independent finite-shot runs per number of dimensions $d$, we demonstrate certified CGLMP violations throughout $d=2$--$8$. For isotropic mixed states tested at a visibility of just 0.05 above the standard-Fourier violation threshold, we likewise observe confidence-certified CGLMP violations throughout the range of $d$ studied. We compare this direct stochastic approach with quantum state tomography, the standard method for characterising an unknown state. In the matched benchmark, CSPSA uses fewer measurement configurations per attempt from $d=6$, whereas tomography requires fewer detected pairs per certified result through $d=8$. We also derive the phase dependence of the CGLMP parameter and clarify the features of its landscape that govern the adaptive search. Because the measurement-setting cost of each CSPSA iteration is independent of dimension, the method offers a particularly attractive route to the certification of high-dimensional entanglement.
\end{abstract}

\maketitle

\section{Introduction}\label{sec:intro}

\blue{Entanglement supplies the nonclassical correlations that fuel many of the advantages pursued in quantum technology, from quantum computation and secure communication to distributed information processing and precision sensing.}
Bell-inequality violation provides a direct certificate of nonclassical correlations, but the measurement settings required to reveal a violation usually depend on the state. This creates a practical problem when transmission, noise, or imperfect preparation changes the relevant basis. Full quantum state tomography (QST) resolves that uncertainty by reconstructing the density matrix, although its measurement and processing requirements increase rapidly with local dimension \cite{qtmtomo_qubits,qtmtomo_qudits,selfguidetomo_CSPSA_noiseresistant_qudits}. A task-specific alternative is to adapt the Bell measurement until a violation is found.

Stochastic optimisation has recently been used to adjust the four local settings of a Clauser--Horne--Shimony--Holt (CHSH) test for unknown two-qubit states \cite{CortesVega2023,CSPSA_CHSH_applyto_fibre}. Complex simultaneous perturbation stochastic approximation (CSPSA) and related methods have also been applied to state estimation, state discrimination, variational entanglement measures and nonlocal witnesses \cite{CSPSA_Alarcon2019,reivew_of_stochastic_optimisation_algo,CSPSA_applyto_statediscrimination,CSPSA_applyto_multiqubit,MatsunagaHo2025}. Previous adaptive Bell searches therefore concentrate on two-outcome tests, whereas high-dimensional stochastic protocols principally reconstruct a state. Here we address a $d$-outcome Bell-test setting in which the Bell statistics align the analyser itself and the adaptively returned setting is validated with fresh finite data.
\blue{The approach is attractive because each update requires only two evaluations of the objective, independently of the number of optimised parameters.}

\blue{High-dimensional entanglement is attractive for quantum communication because it provides a larger outcome alphabet and can improve noise tolerance in specific protocols \cite{Dada2011_exp_qudits_CGLMP,Durt_qutrits_theory_maxnoise}.}
Here we develop and numerically evaluate such a self-aligning high-dimensional Bell-certification procedure using the CGLMP inequality \cite{CGLMP}. High-dimensional CGLMP violations have been measured with photonic states up to $d=12$ and $16$ using prescribed analysers \cite{Dada2011_exp_qudits_CGLMP,Lo2016_exp_qudits_CGLMP}. We instead keep the CGLMP coefficients and local upper bound fixed while the measured signed Bell value adjusts the four offsets defining Alice's and Bob's Fourier projective bases. The Bell test performs its own phase alignment and terminates in a fresh finite-shot certificate without the need for density-matrix reconstruction.

Each CSPSA update acquires the CGLMP value at two simultaneous perturbations, $S_{+}$ and $S_{-}$, and uses their difference to update every phase. \blue{After the final update, we acquire four new joint-outcome tables at the returned phases. These fresh data give the reported Bell value. We then subtract a conservative statistical margin, calculated from Hoeffding's inequality at a confidence level chosen before the data are examined \cite{Hoeffding1963}. The result is a one-sided lower confidence bound, and we call the run certified only if it remains greater than the classical limit $S_d=2$.} Signed physical-multinomial simulations with $100$ independent trials at each number of dimensions, $d$, produce confidence-certified violations throughout $d=2$--$8$: the certified counts at the main operating point are $100$, $74$, $84$, $80$, $80$, $78$ and $87$, respectively, in one independently seeded ensemble of 100 complete runs per dimension. The separation between the narrow terminal sampling error and the broad between-run distribution identifies entry into different stationary basins, rather than shot noise, as the principal limit on reliability.

\blue{ CSPSA varies four offsets within Alice's and Bob's Fourier projective bases, while the unknown state may be pure or mixed. 
We use an affine Schmidt-phase ramp as a controlled example of an unknown Bell-test misalignment and show that adapting the measurement offsets identifies settings that recover the available Bell violation. 
In such a ramp, the phase changes by the same amount from one mode to the next, so a common shift of Alice's two offsets cancels it. Independent Schmidt phases provide a more demanding test because they contain additional structure that four offsets cannot generally compensate as $d$ grows. Together with the mixed-state study below, these cases probe the reach of a Bell test in which only the four Fourier measurement offsets are adapted.
}

We test mixed-state performance using isotropic white-noise mixtures, often called Werner-type states. At the standard Fourier setting, their exact Bell value scales linearly with visibility. Fresh measurements taken after the final CSPSA update determine whether this Bell margin is sufficient to certify a violation with finite data. 
 The optimisation also exposes a second question: what, precisely, is being optimised? 
 Our analysis separates two different optimisation problems. For a fixed maximally entangled input, we derive an explicit formula $F_d(\bm\theta)$ for the signed CGLMP value as a function of the four Fourier measurement offsets \(\boldsymbol\theta\).  Maximising \(F_d\) changes the measurements varied by CSPSA while keeping the state fixed. We compare this with the Ac\'{\i}n-type quantity $G_d(\bm\theta)$, which is the largest eigenvalue of the Bell operator ~\cite{Acin_QtmNonlocality}. \blue{For $d=2$--$6$, the gradient and Hessian establish the Fourier point as a strict local maximum, while searches across the full phase range provide numerical evidence that no higher value occurs within the four-offset measurement family.}

\blue{The comparison is task-specific: QST reconstructs the full state, uses that reconstruction to select suitable CGLMP measurement settings, and then performs the Bell test, whereas CSPSA uses the measured CGLMP value itself to search directly for settings that reveal a violation. This distinction may be especially relevant to entanglement-based QKD protocols such as E91, where channel-induced drift can alter the effective measurement alignment during operation and a direct adaptive Bell-objective search could help recover settings that reveal a violation without repeatedly reconstructing the full state.}
Our resource comparison separates settings from detected pairs. With the number of iterations $K=100$, $N_{\mathrm{opt}}=1000$ pairs per setting pair and a fresh $N_{\mathrm{cert}}=10000$-pair terminal acquisition, the adaptive route uses $804$ joint-setting configurations and $840000$ detected pairs per attempt. Its raw configuration count first falls below that of the implemented QST scheme at $d=6$. After accounting for the probability of successful certification, the $K=100$ point estimate first favours CSPSA at $d=8$, although the transformed confidence intervals overlap. In the reported $d=5$--$8$, $K=50$ sensitivity scan, the point crossover occurs at $d=6$ and the transformed intervals separate from $d=7$. No $K=100$ detected-pair crossover is observed through $d=8$. The result is a direct, statistically valid Bell certificate with dimension-independent per-iteration setting complexity and an operationally explicit tradeoff between reconfiguration and pair acquisition.

\vspace{20pt}

\section{CGLMP phase measurements and adaptive certification}\label{sec:method}

\subsection{Fixed Bell functional and projective realisation}

In the $(2,2,d)$ Bell scenario, Alice and Bob choose settings $a,b\in\{0,1\}$ and obtain outcomes $r,s\in\{0,\ldots,d-1\}$. The CGLMP parameter is the fixed linear combination
\begin{widetext}
\begin{equation}
\begin{split}
S_d=\sum_{k=0}^{[d/2]-1}\left(1-\tfrac{2k}{d-1}\right)\{&+[P(A_0=B_0+k)+P(B_0=A_1+k+1)+P(A_1=B_1+k)+P(B_1=A_0+k)]\\
&-[P(A_0=B_0-k-1)+P(B_0=A_1-k)+P(A_1=B_1-k-1)+P(B_1=A_0-k-1)]\},
\end{split}
\label{eq:CGLMP Parameter}
\end{equation}
\end{widetext}
where $[x]$ denotes the integer part of $x$ and all outcome equalities are modulo $d$. Every local-hidden-variable model obeys $S_d\leq2$. The lower local bound is asymmetric for $d>2$, so a violation always means the signed condition $S_d>2$; while $|S_d|>2$ remains valid for $d=2$ violations.

For example, the modular event probabilities are obtained directly from the joint table,
\begin{equation}
P(A_a=B_b+q)=\sum_{s=0}^{d-1}p(s+q,s|a,b).
\end{equation}
The coefficients in Eq.~\eqref{eq:CGLMP Parameter}, including the local upper bound, remain unchanged throughout the optimisation.

The maximally entangled reference state is
\begin{equation}
|\Psi_d\rangle=\frac{1}{\sqrt d}\sum_{j=0}^{d-1}|j\rangle_A\otimes|j\rangle_B.
\label{eq:max ent corr state}
\end{equation}
The standard CGLMP measurements are Fourier bases with phase offsets  $\bm\theta_\star = (\alpha_0,\alpha_1,\beta_0,\beta_1)=(0,1/2,1/4,-1/4)$. We retain this \blue{experimentally accessible} family while allowing all four offsets to vary:
\begin{align}
|a,r\rangle_A&=\frac{1}{\sqrt d}\sum_{j=0}^{d-1}
\exp\!\left[\frac{2\pi i j}{d}(r+\alpha_a)\right]|j\rangle,
\label{eq:alicebasis}\\
|b,s\rangle_B&=\frac{1}{\sqrt d}\sum_{j=0}^{d-1}
\exp\!\left[\frac{2\pi i j}{d}(-s+\beta_b)\right]|j\rangle.
\label{eq:bobbasis}
\end{align}
For real offsets, the projectors $M^A_{r|a}=|a,r\rangle\langle a,r|$ and $M^B_{s|b}=|b,s\rangle\langle b,s|$ are orthogonal and complete. The joint probabilities are
\begin{equation}
p(r,s|a,b)=\mathrm{Tr}\!\left[\rho\left(M^A_{r|a}\otimes M^B_{s|b}\right)\right],
\label{eq:bornprob}
\end{equation}
with $\sum_{r,s}p(r,s|a,b)=1$ for every setting pair.

For $\bm\theta=(\alpha_0,\alpha_1,\beta_0,\beta_1)$, the associated Bell operator is
\begin{equation}
\mathcal B_d(\bm\theta)=\sum_{a,b=0}^{1}\sum_{r,s=0}^{d-1}
c^{(d)}_{abrs}M^A_{r|a}(\bm\theta)\otimes M^B_{s|b}(\bm\theta),
\label{eq:belloperator}
\end{equation}
where $c^{(d)}_{abrs}$ are precisely the coefficients in Eq.~\eqref{eq:CGLMP Parameter}. Hence $S_d(\rho,\bm\theta)=\mathrm{Tr}[\rho\mathcal B_d(\bm\theta)]$. We note here that changing $\bm\theta$ changes the projective realisation and the operator, but it does not define a new Bell functional.

\subsection{CSPSA update and physical estimator}

CSPSA estimates a gradient from two simultaneous perturbations of all parameters \cite{CSPSA_Alarcon2019,Spall_SPSA}. Although the general method accommodates complex variables, Eqs.~\eqref{eq:alicebasis} and \eqref{eq:bobbasis} contain four real offsets. At iteration \(k\), \(\bm{\Delta}_k\) is a four-component perturbation vector, and each component is independently chosen to be either \(-1\) or \(+1\). It determines whether each of the four measurement-phase offsets is shifted down or up during that CSPSA step. Thus, the random vector $\bm\Delta_k\in\{-1,+1\}^4$ defines
\begin{equation}
\bm\theta_{k\pm}=\bm\theta_k\pm c_k\bm\Delta_k.
\end{equation}

The two measured CGLMP estimates $S_{d,k}^{\pm}$ give
\begin{equation}
\widetilde g_{k,i}=\frac{S_{d,k}^{+}-S_{d,k}^{-}}
{2c_k\Delta_{k,i}},
\label{eq:gradient}
\end{equation}
and the phases are updated to maximise the signed Bell value,
\begin{equation}
\bm\theta_{k+1}=\bm\theta_k+a_k\widetilde{\bm g}_k.
\label{eq:update}
\end{equation}
The gain sequences are
\begin{equation}
a_k=\frac{a}{(k+1+A)^s},\qquad c_k=\frac{b}{(k+1)^r}.
\label{eq:gains}
\end{equation}

\begin{table}[t]
\centering
\setlength{\tabcolsep}{6.4pt}
\begin{tabular}{c|ccccc}
\toprule
$d$ & $a$ & $A$ & $s$ & $b$ & $r$\\
\midrule
2 & 1 & 5  & 0.8  & 0.8 & 0.25\\
3 & 1 & 10 & 0.6  & 0.8 & 0.19\\
4 & 1 & 15 & 0.5  & 0.8 & 0.19\\
5 & 1 & 25 & 0.55 & 0.8 & 0.19\\
6 & 1 & 30 & 0.55 & 0.8 & 0.19\\
7 & 1 & 35 & 0.55 & 0.8 & 0.19\\
8 & 1 & 40 & 0.55 & 0.8 & 0.19\\
\bottomrule
\end{tabular}
\caption{Gain parameters used in every reported ensemble. \blue{For $d>4$ we use $A=5d$ and $s=0.55$, with the other parameters fixed.} All gains were fixed before and held unchanged across the reported comparisons.}
\label{tab:gains}
\end{table}

One Bell evaluation comprises the four joint settings $(a,b)$. For each setting pair, $N$ detected qudit pairs are recorded in a single multinomial table $n^{ab}_{rs}$, with $\sum_{rs}n^{ab}_{rs}=N$ and $\widetilde p(r,s|a,b)=n^{ab}_{rs}/N$. Writing the CGLMP coefficients as $w^{(d)}_{abrs}$, the physical estimator is
\begin{equation}
\widetilde S_d=\sum_{abrs}w^{(d)}_{abrs}\widetilde p(r,s|a,b).
\label{eq:physical-estimator}
\end{equation}
The four tables preserve the covariance between all modular events derived from the same setting. The plug-in standard error is
\begin{equation}
\widetilde{\sigma}_{S}^{2}=\sum_{a,b}\frac{1}{N-1}
\left[\sum_{r,s}w_{abrs}^{2}\widetilde p_{abrs}
-\left(\sum_{r,s}w_{abrs}\widetilde p_{abrs}\right)^2\right].
\label{eq:bell-se}
\end{equation}

Only $S^+_{d,k}$ and $S^-_{d,k}$ are acquired during an update. Their midpoint $(S^+_{d,k}+S^-_{d,k})/2$ is a convergence proxy formed from required data. \blue{Exact values at $\bm\theta_k$ are evaluated only as simulation diagnostics and are not treated as measurements.} After $K$ updates, four newly acquired multinomial tables at $\bm\theta_K$ give $\widetilde S_{d,\mathrm{term}}$. Collecting these counts independently of the data used to select $\bm\theta_K$ preserves the validity of the terminal confidence bound following adaptive setting selection. 
For a fixed pair of measurement settings $(a,b)$, each possible joint outcome $(r,s)$ contributes a coefficient $w^{(d)}_{abrs}$ to the Bell estimate. To quantify how much this contribution can vary across the possible outcomes, we take the difference between the largest and smallest coefficient for that setting pair. That is,
\begin{equation}
R_{ab}
=
\max_{r,s} w^{(d)}_{abrs}
-
\min_{r,s} w^{(d)}_{abrs}.
\end{equation}
The distribution-free one-sided $(1-\alpha)$ Hoeffding lower bound used here is \cite{Hoeffding1963}
\begin{equation}
S_d^{\mathrm L}=\widetilde S_{d,\mathrm{term}}-
\sqrt{\frac{\ln(1/\alpha)}{2}\sum_{a,b}\frac{R_{ab}^2}{N_{\mathrm{cert}}}}.
\label{eq:hoeffding}
\end{equation}

Here, $\widetilde S_{d,\mathrm{term}}$ is the signed CGLMP value obtained from four fresh terminal joint-outcome tables, each containing $N_{\mathrm{cert}}$ detected pairs, and $R_{ab}$ is the range of the possible single-pair CGLMP contributions for setting pair $(a,b)$. The square-root term is a conservative statistical margin that increases with the required confidence and decreases as $N_{\mathrm{cert}}^{-1/2}$. Here, $\alpha=0.05$ is the predeclared upper bound on the probability that statistical fluctuation causes $S_d^{\mathrm L}$ to exceed the underlying mean Bell value; equivalently, $1-\alpha=0.95$ is the one-sided confidence level used throughout. 
Because the terminal data are independent of those used during optimisation, $S_d^{\mathrm L}$ provides a one-sided lower confidence bound on the Bell value at the returned settings without requiring a Gaussian approximation.  

A trial is called certified only when the predeclared condition $S_d^{\mathrm L}>2$ holds. Across $L$ independent trials, certified-success probabilities are accompanied by two-sided $95\%$ Wilson intervals \cite{Wilson1927}.

Each update uses two Bell evaluations, hence eight joint-setting configurations. Including the fresh terminal test, the operational costs are
\begin{align}
C_{\mathrm{set}}&=8K+4,\label{eq:config-resource}\\
C_{\mathrm{pair}}&=8KN_{\mathrm{opt}}+4N_{\mathrm{cert}}.
\label{eq:pair-resource}
\end{align}
For the main operating point $K=100$, $N_{\mathrm{opt}}=1000$ and $N_{\mathrm{cert}}=10000$, Eqs.~\eqref{eq:config-resource} and \eqref{eq:pair-resource} give $804$ configurations and $840000$ detected pairs per attempt.

Algorithm~\ref{alg:phase-only-cspsa} summarises the complete operational loop. In particular, it separates the two adaptively used perturbation acquisitions from the fresh terminal acquisition and never selects a result by its magnitude.

\begin{algorithm}[t]
\small
\DontPrintSemicolon
\caption{High-dimensional CSPSA optimisation and fresh CGLMP certification}
\label{alg:phase-only-cspsa}
\KwIn{$d$, fixed $K,N_{\mathrm{opt}},N_{\mathrm{cert}}$, gains $(a,A,s,b,r)$, confidence level $1-\alpha$}
\KwOut{Signed terminal estimate $\widetilde S_{d,\mathrm{term}}$, lower bound $S_d^{\mathrm L}$ and the decision $S_d^{\mathrm L}>2$}
Draw $\bm\theta_0\in[0,1)^4$;
\For{$k=0,\ldots,K-1$}{
  Draw $\bm\Delta_k\in\{-1,+1\}^4$ and evaluate $a_k,c_k$ from Eq.~\eqref{eq:gains};
  Form $\bm\theta_{k\pm}=\bm\theta_k\pm c_k\bm\Delta_k$;
  \For{$\sigma\in\{+,-\}$}{
    Acquire four $d\times d$ multinomial tables at $\bm\theta_{k\sigma}$, using $N_{\mathrm{opt}}$ detected pairs per joint setting;
    Compute the signed Bell estimate $S_{d,k}^{\sigma}$ from Eq.~\eqref{eq:physical-estimator};
  }
  Optionally store $(S_{d,k}^{+}+S_{d,k}^{-})/2$ as a no-cost convergence proxy;
  Form $\widetilde{\bm g}_k$ from Eq.~\eqref{eq:gradient} and update $\bm\theta_{k+1}$ by Eq.~\eqref{eq:update};
}
At the predeclared $K$, acquire four \emph{new} tables at $\bm\theta_K$, using $N_{\mathrm{cert}}$ detected pairs per setting;
  Compute the signed terminal estimate from Eq.~\eqref{eq:physical-estimator}, its standard error from Eq.~\eqref{eq:bell-se}, and the independent lower bound in Eq.~\eqref{eq:hoeffding};
\Return{certified if and only if $S_d^{\mathrm L}>2$};
\end{algorithm}

\begin{table*}[t!]
\centering
\setlength{\tabcolsep}{8pt}
\begin{tabular}{c@{\hspace{1em}}cccc}
\toprule
$d$ & Median $\widetilde S_{d,\mathrm{term}}$ & Mean $\widetilde S_{d,\mathrm{term}}$
& Mean terminal SE & Hoeffding certified, $n/100$ [Wilson $95\%$]\\
\midrule
2 & 2.8295 & 2.8271 & 0.0141 & 100 $[96.3,100.0]\%$\\
3 & 2.8664 & 2.4715 & 0.0131 & 74 $[64.6,81.6]\%$\\
4 & 2.8825 & 2.6354 & 0.0119 & 84 $[75.6,89.9]\%$\\
5 & 2.9067 & 2.6122 & 0.0115 & 80 $[71.1,86.7]\%$\\
6 & 2.9147 & 2.6270 & 0.0115 & 80 $[71.1,86.7]\%$\\
7 & 2.9219 & 2.6069 & 0.0114 & 78 $[68.9,85.0]\%$\\
8 & 2.9316 & 2.7439 & 0.0119 & 87 $[79.0,92.2]\%$\\
\bottomrule
\end{tabular}
\caption{Final Bell-test certification of maximally entangled inputs. Each row contains $100$ trials with $K=100$, $N_{\mathrm{opt}}=1000$ and $N_{\mathrm{cert}}=10000$. Sampled, exact-at-returned-settings and Hoeffding-certified success counts coincide at this operating point.}
\label{tab:terminal}
\end{table*}

\subsection{Input families and simulation design}

We consider the Schmidt-correlated maximally entangled states
\begin{equation}
|\Psi_{d,\bm\phi}\rangle=\frac{1}{\sqrt d}\sum_{j=0}^{d-1}e^{i\phi_j}|j,j\rangle.
\label{eq:phase-state}
\end{equation}
The supported affine family has
\begin{equation}
\phi_j=\phi_0+\frac{2\pi\delta j}{d}\pmod{2\pi}.
\label{eq:affine-ramp}
\end{equation}
Successive Schmidt-mode phases therefore differ by the same constant increment $2\pi\delta/d$; $\phi_0$ is an irrelevant global phase. This is what we mean by an affine phase ramp in the present study.
\blue{For each complete optimisation run, we choose a new slope $\delta$ uniformly at random from $0\leq\delta<d$. The resulting phase ramp is then kept fixed throughout that run, and its value is not given to CSPSA.}
It is compared with independent phases satisfying $0\leq\phi_j<2\pi$, after fixing the irrelevant global phase. The departure from an affine ramp is quantified for $d>2$ by
\begin{equation}
R_{\phi}=\left\{\frac{1}{d-2}\sum_{j=2}^{d-1}
\left[\arg e^{i(\phi_j-\phi_0-j(\phi_1-\phi_0))}\right]^2\right\}^{1/2},
\label{eq:phase-residual}
\end{equation}
and $R_{\phi}=0$ for $d=2$. Equation~\eqref{eq:affine-ramp} has $R_\phi=0$ and can be absorbed by a common shift of Alice's two phase offsets.

Noise tolerance is tested with the isotropic family
\begin{equation}
\rho_v=v|\Psi_d\rangle\langle\Psi_d|+(1-v)\frac{\mathbb I_{d^2}}{d^2}.
\label{eq:isotropic}
\end{equation}
Here $v\in[0,1]$ is the visibility, namely the weight of the maximally entangled component, and $d$ is the Hilbert-space dimension of each local subsystem. 
Since the CGLMP Bell operator is traceless, $\operatorname{Tr}[\mathcal B_d(\bm\theta)]=0$, the maximally mixed component contributes no Bell signal:
\[
S_d(v,\bm\theta)
=\operatorname{Tr}[\rho_v\mathcal B_d(\bm\theta)]
=vF_d(\bm\theta).
\]
 We therefore define the standard-Fourier reference visibility
\begin{equation}
v_{\mathrm F}=\frac{2}{F_d(\bm\theta_\star)}.
\label{eq:fourier-visibility}
\end{equation}
At $v=v_{\mathrm F}$ the exact value at the standard Fourier measurement setting, $\bm\theta_\star$, is the local bound. We note that, although $v_{\mathrm F}$ marks the visibility at which the standard Fourier measurement reaches the local bound, other allowed measurement settings may still violate the Bell inequality at lower visibilities.

Unless stated otherwise, every ensemble contains $L=100$ independently seeded trajectories. That is, each complete run uses an independent random 
stream for its initial phases, CSPSA perturbations and simulated measurement records; input-state parameters are also redrawn when the declared ensemble samples over states. 
The main finite-shot results use $K=100$, $N_{\mathrm{opt}}=1000$ and $N_{\mathrm{cert}}=10000$. The initial phase vector \(\bm\theta_0=(\theta_{0,1},\ldots,\theta_{0,4})\) is generated by drawing each of its four components independently and uniformly from the interval \([0,1)\), 
and every perturbation component is independently selected from $\{-1,+1\}$. Means and standard deviations describe the full between-run distribution; medians and interquartile ranges describe its centre and middle half. The terminal standard error in Eq.~\eqref{eq:bell-se} is also reported separately from both.

\section{Signed finite-shot performance through \texorpdfstring{$d=8$}{d=8}}\label{sec:performance}

Table~\ref{tab:terminal} summarizes the Bell values and confidence-certification outcomes obtained from the new terminal datasets collected after optimization. Every qubit run certifies a violation, while, at higher dimensions, the certified rate is $74$--$87\%$ throughout $d=3$--$8$ under one fixed algorithm and photon budget. The median remains close to the standard Fourier reference while the mean is pulled down by lower stationary outcomes, indicating that most trajectories reach the expected high-value region. Individual terminal Bell estimates are precise, with mean standard errors of only $0.011$--$0.013$. The between-run standard deviation of $0.49$--$0.68$ reflects convergence to different optimisation basins and points to improved basin entry as the natural route to still higher certification rates.

The signed trajectories explain the mean--median separation. At $N_{\mathrm{opt}}=1000$, the upper-branch fractions for $d=2$--$8$ are $1.00$, $0.74$, $0.84$, $0.80$, $0.80$, $0.78$ and $0.87$. The lower-branch means for $d=3$--$8$ lie between $1.31$ and $1.49$, well below the local upper bound but far from the statistical width of an upper-branch terminal estimate. At each update, the midpoint $(S^+_{d,k}+S^-_{d,k})/2$ provides an estimate of the Bell value near the current setting using the two measurements already required by CSPSA without using up any additional photons.   
Among runs for which this midpoint crosses the local bound, the median first crossing occurs at iteration $13$ for $d=2$, rises to $25$ for $d=7$, and is $24$ for $d=8$. Thus, runs that reach the violating branch generally do so well before the $100$-iteration limit. The unsuccessful runs are therefore not merely progressing more slowly towards the same solution; the longer iteration study shows that most instead remain in a different, lower-valued optimisation basin.

\begin{figure*}[t]
\centering
\includegraphics[width=0.98\textwidth]{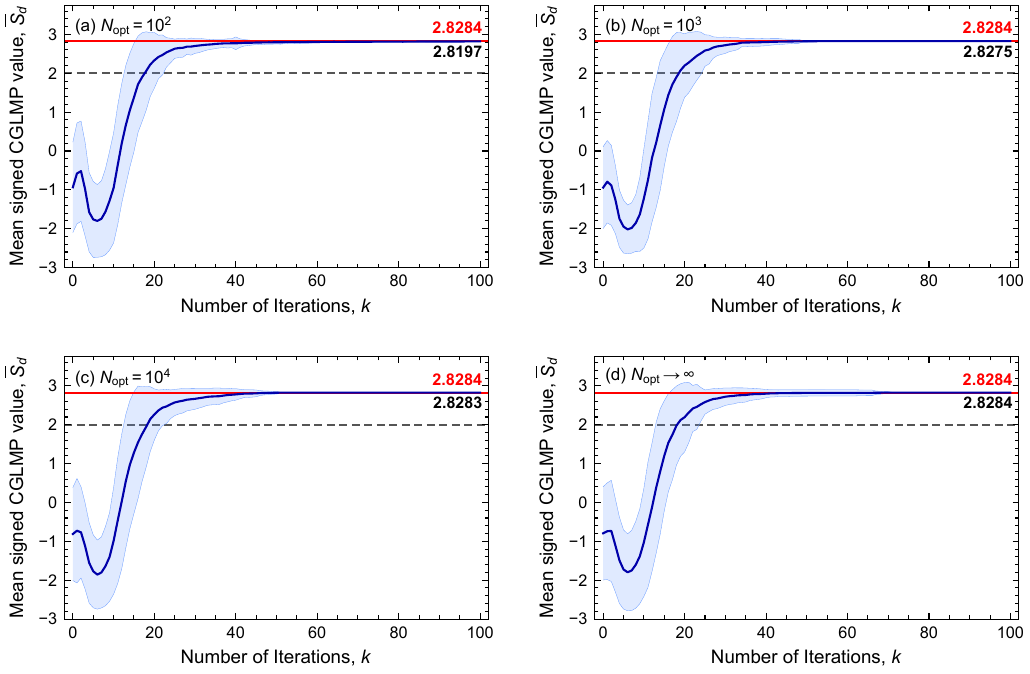}
\caption{Per-update CSPSA convergence for a maximally entangled qubit pair at four optimisation photon budgets: (a) $N_{\mathrm{opt}}=100$, (b) $N_{\mathrm{opt}}=1000$, (c) $N_{\mathrm{opt}}=10000$ detected pairs per joint setting, and (d) the noiseless $N_{\mathrm{opt}}\to\infty$ reference. Each panel contains $100$ independently seeded $K=100$ trajectories. At each iteration, the dark-blue curve is the mean of the $100$ exact central simulation values $S_2(\bm\theta_k)$. The pale-blue band shows how those values vary around the mean: its lower and upper edges, marked by fine blue lines, lie one between-run sample standard deviation below and above the mean. It is therefore not an estimator error bar or a confidence interval for one Bell measurement. Each perturbed Bell evaluation uses the indicated $N_{\mathrm{opt}}$ detected pairs per joint setting; $N_{\mathrm{opt}}\to\infty$ means that the perturbed Bell values are calculated from exact Born probabilities rather than sampled counts. A band can nevertheless remain at a finite iteration because the independently seeded runs have different initial phases and perturbation sequences. The grey dashed line is the local bound $S_2=2$, the red line and number give the exact standard-Fourier fixed-state reference, and the black number gives the final ensemble mean. The central curves are simulation diagnostics \blue{evaluated along the optimisation trajectories} and require no additional photon sampling; final Bell-test certification uses four newly acquired joint-outcome tables at the returned setting.}
\label{fig:d2-budget-convergence}
\end{figure*}

Our plots in Figure~\ref{fig:d2-budget-convergence} emphasise the motion of the optimiser at every update at $d=2$ for three finite optimisation photon allocations, \(N_{\mathrm{opt}}=100,1000,\) and \(10\,000\) detected pairs per joint setting pair, together with the noiseless \(N_{\mathrm{opt}}\to\infty\) reference. \blue{For each iterate $\bm\theta_k$} we evaluate the exact objective $S_2(\bm\theta_k)$ in simulation, so the resulting $101$-point trace shows the central trajectory from the initial setting through update $K=100$. We note that this does not represent an additional central finite-shot acquisition, since the physical data used by the update remain the two perturbed values $S^+_{d,k}$ and $S^-_{d,k}$, and certification is performed using four newly acquired joint-outcome count tables at the final setting \(\bm\theta_K\); none of the data used to choose that setting are reused for certification. 
 The dark-blue line is the mean Bell value across the \(L=100\) runs at each iteration. The pale-blue band shows the spread of those 100 values around the mean: its lower and upper edges lie one between-run sample standard deviation below and above the mean, respectively, and the fine blue lines mark these two edges. The band therefore describes variation between complete runs rather than a confidence interval or an additional Bell measurement.
 
\begin{figure*}[t]
\centering
\includegraphics[width=0.98\textwidth]{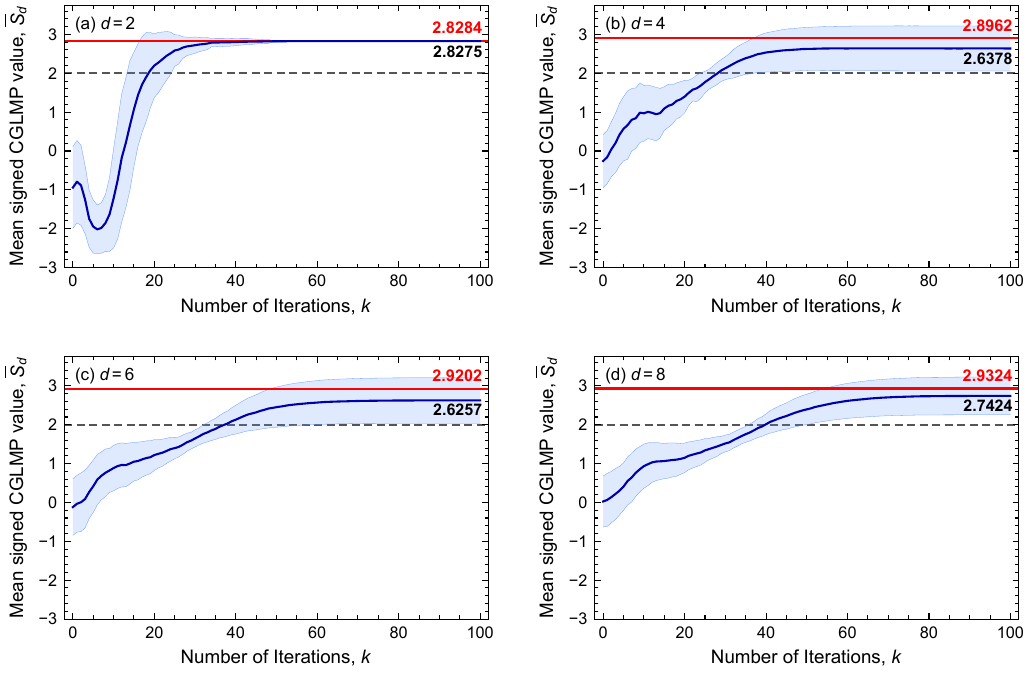}
\caption{Per-update phase-only CSPSA convergence for maximally entangled inputs at $N_{\mathrm{opt}}=1000$ detected pairs per joint setting: (a) $d=2$, (b) $d=4$, (c) $d=6$, and (d) $d=8$. Each panel contains $100$ independently seeded trajectories from the initial setting through $K=100$. At each iteration, the dark-blue curve is the mean of the $100$ exact signed central simulation values $S_d(\bm\theta_k)$. The pale-blue again band shows how those values vary around the mean: its lower and upper edges, marked by fine blue lines, lie one between-run sample standard deviation below and above the mean. It is therefore not an estimator error bar or a confidence interval for one Bell measurement. The grey dashed line is the local bound $S_d=2$, the red line and number give the standard-Fourier fixed-state reference $F_d(\bm\theta_\star)$, and the black number gives the final ensemble mean. The reference is not asserted to be a globally certified maximum over the four-offset Fourier measurement family for $d>2$. The central values are simulation diagnostics \blue{evaluated at each iterate} and require no additional photon sampling. The optimiser uses the two perturbed finite-shot Bell evaluations, while final Bell-test certification uses four newly acquired joint-outcome tables at the returned setting.}
\label{fig:per-update-convergence}
\end{figure*}

Figure~\ref{fig:per-update-convergence} shows similar per-update plots for $d=2,4,6,8$ at the main optimisation budget $N_{\mathrm{opt}}=1000$.  At $d=4,6,8$, successful trajectories approach the standard-Fourier neighbourhood, while lower stationary branches pull the ensemble mean below the red reference and keep the between-run standard deviation finite. Thus it is clear the broad band at late iterations is not residual uncertainty in a single Bell measurement. Rather, it records the coexistence of optimisation basins across the $100$ independent trials. Table~\ref{tab:terminal} reports the median alongside the mean as a complementary summary.

\begin{figure*}[t]
\centering
\includegraphics[width=0.96\textwidth]{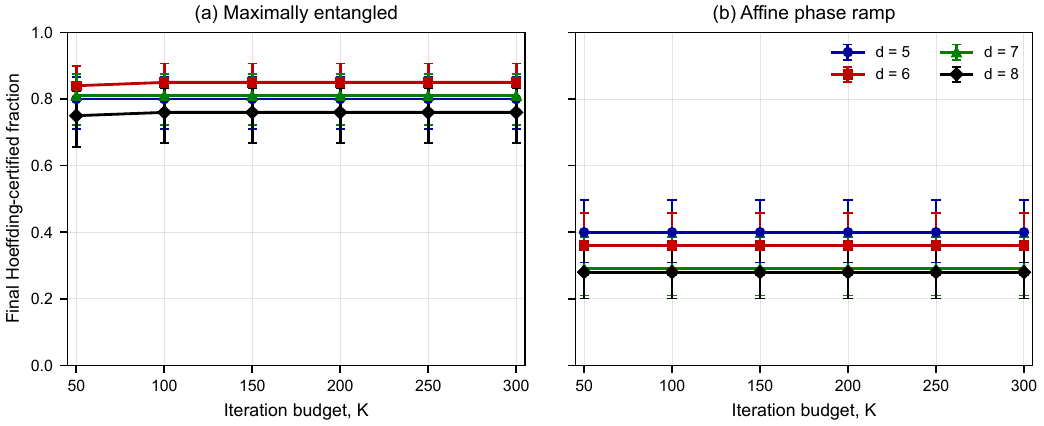}
\caption{Final Bell-test certification rates at alternative iteration budgets for the high-dimensional scan $d=5,\ldots,8$. Panel (a) shows maximally entangled inputs, and panel (b) shows maximally entangled inputs with a randomly chosen affine phase ramp. Colour and marker jointly identify the dimension. Points show the observed fraction from $100$ trajectories and bars are pointwise Wilson $95\%$ intervals. Each $K\in\{50,100,150,200,250,300\}$ is costed as an alternative terminal design with a fresh $N_{\mathrm{cert}}=10000$ acquisition; certificate data are not accumulated across $K$. Because successive values use prefixes of the same trajectories, values at different $K$ within a panel are correlated. The near-constant rates show that additional updates refine trajectories already in the upper basin rather than materially changing basin occupancy.}
\label{fig:high-d-certification}
\end{figure*}

A separate, independently seeded high-dimensional iteration-budget ensemble distinguishes convergence within a basin from the probability of entering it. For maximally entangled inputs, the confidence-certified rates at $K=50$ are $80\%$, $84\%$, $81\%$ and $75\%$ for $d=5$--$8$; at $K=100$ they are $80\%$, $85\%$, $81\%$ and $76\%$. Increasing the tested budget further to $K=300$ changes none of those counts: the only changes across the full comparison are the single additional successes at $d=6$ and $8$, both already present at $K=100$. The upper-branch median improves markedly between $K=50$ and $100$, but longer runs do not materially alter branch occupancy. This is why we have chosen $K=100$ as the main operating point. Figure~\ref{fig:high-d-certification} displays the corresponding final certification rates 
and their pointwise Wilson intervals.
For the affine-ramp test case, allowing more optimisation iterations did not improve the certification rate in dimensions \(d=5\)–\(8\). The rates remained \(40\%\), \(36\%\), \(29\%\), and \(28\%\), respectively, whether the algorithm was run for \(K=50\) iterations or \(K=300\). Thus, the unsuccessful runs were not simply needing more iterations to converge.  
Although the settings at iteration \(K\) are informed by all preceding optimisation updates, data from different values of \(K\) are not combined and each value of \(K\) represents a separate choice of when to stop the optimisation and perform the final certification test. Because increasing \(K\) did not improve the certification rate, we found that stopping at \(K=50\) required the fewest detected pairs in every dimension. Further improvement to the algorithm should therefore focus on helping more runs reach the high-valued solution, for example, through better initial settings, a clearly specified restart strategy, or a broader family of measurements. 

\begin{table*}[t]
\centering
\setlength{\tabcolsep}{5.2pt}
\begin{tabular}{c@{\hspace{1.1em}}ccc}
\toprule
$d$
& \shortstack{Maximally entangled\\$K=50$}
& \shortstack{Maximally entangled\\$K=100$}
& \shortstack{Maximally entangled with randomly\\chosen affine phase ramp\\$K=50,\ldots,300$}\\
\midrule
5 & 80\% & 80\% & 40\%\\
6 & 84\% & 85\% & 36\%\\
7 & 81\% & 81\% & 29\%\\
8 & 75\% & 76\% & 28\%\\
\bottomrule
\end{tabular}
\caption{Final Bell-test certification rates in the high-dimensional iteration-budget comparison. Maximally entangled and affine-ramp columns use separate ensembles. Each value of $K$ is treated as a separate stopping point, and its photon cost is calculated separately.}
\label{tab:iteration-scan}
\end{table*}

\begin{figure*}[t]
\centering
\includegraphics[width=0.96\textwidth]{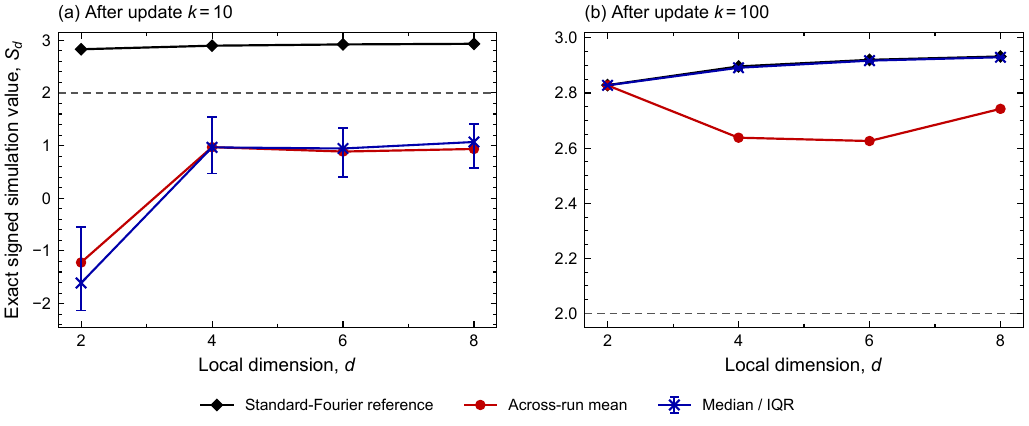}
\caption{CSPSA attainment of the standard-Fourier fixed-state reference in four representative dimensions. Panels show the same $100$ independently seeded $N_{\mathrm{opt}}=1000$ trajectories after (a) update $k=10$ and (b) update $k=100$. Red circles give the arithmetic mean over the $100$ independent optimisation runs, blue crosses give the median with between-run IQR, black diamonds give $F_d(\bm\theta_\star)$, and the grey dashed line is the local bound $S_d=2$. The plotted ordinate is exact signed simulation truth evaluated at \blue{the iterate $\bm\theta_k$}; it is a diagnostic that consumes no additional detected pairs and is not the final Bell-test certificate. The same trajectories appear in both panels. }
\label{fig:reference-attainment}
\end{figure*}

Figure~\ref{fig:reference-attainment} condenses \blue{the same trajectories} into an early-versus-final comparison. At $k=10$ the four representative dimensions remain far from the standard-Fourier reference. By $k=100$, however, every median lies within $0.17\%$ of that reference. The all-run means at $d=4,6,8$ remain lower because $16\%$, $20\%$ and $13\%$ of the respective ensembles occupy the lower stationary branch. Thus the median records the typical upper-basin trajectory, whereas the mean retains the operationally important cost of failed basin entry.

\section{Scope of the four-parameter Fourier measurements and noise tolerance}\label{sec:scope}

Table~\ref{tab:scope} compares two classes of maximally entangled input states: those with the same phase increment between successive modes and those with independently chosen mode phases.
 For the affine-ramp inputs, the residual $R_\phi$, which measures departure from a constant phase increment, is zero at $d=2$ and approximately $10^{-15}$ for $d=3$--$8$. The small nonzero values are caused only by numerical rounding. Their upper-branch runs reach the same standard-Fourier reference value as the zero-ramp input. Generic independent phases have median residuals between $1.44$ and $1.91$ radians for $d=3$--$8$, and their certified rate falls rapidly with dimension. At $d=2$, every diagonal relative phase is affine up to a global phase, so the two classes coincide.

\begin{table*}[t]
\centering
\setlength{\tabcolsep}{7.5pt}
\begin{tabular}{c@{\hspace{1.2em}}cc@{\hspace{1.2em}}c}
\toprule
$d$ & Affine-ramp certified [Wilson $95\%$] & Arbitrary-phase certified [Wilson $95\%$]
& Arbitrary-phase median $R_\phi$\\
\midrule
2 & 100 $[96.3,100.0]\%$ & 100 $[96.3,100.0]\%$ & 0\\
3 & 70 $[60.4,78.1]\%$ & 47 $[37.5,56.7]\%$ & 1.444\\
4 & 56 $[46.2,65.3]\%$ & 25 $[17.5,34.3]\%$ & 1.911\\
5 & 40 $[30.9,49.8]\%$ & 10 $[5.5,17.4]\%$ & 1.670\\
6 & 36 $[27.3,45.8]\%$ & 3 $[1.0,8.5]\%$ & 1.771\\
7 & 29 $[21.0,38.5]\%$ & 3 $[1.0,8.5]\%$ & 1.784\\
8 & 28 $[20.1,37.5]\%$ & 0 $[0.0,3.7]\%$ & 1.742\\
\bottomrule
\end{tabular}
\caption{Scope of the four-offset Fourier measurement family. Each entry uses $100$ trials with the same fresh finite-shot terminal test. Independent phases provide a quantitative boundary test for this measurement parametrisation.}
\label{tab:scope}
\end{table*}

\begin{figure*}[t]
\centering
\includegraphics[width=0.94\textwidth]{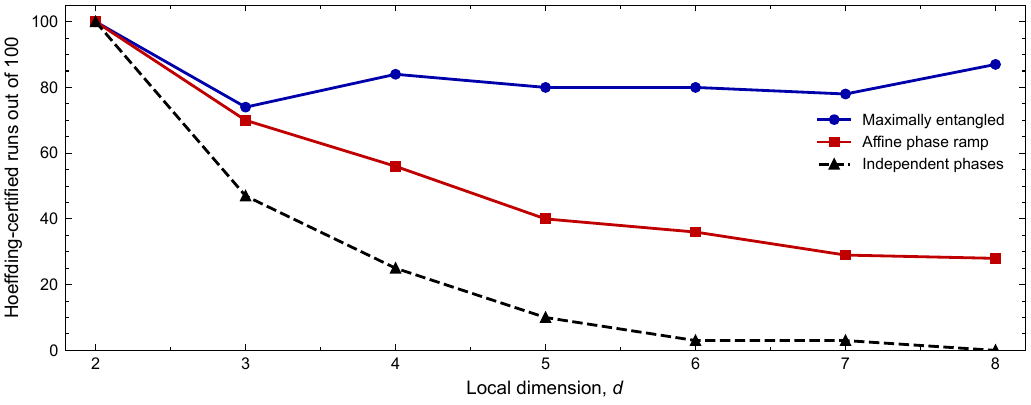}
\caption{Final Bell-test certification across the three studied input families. Points reproduce Tables~\ref{tab:terminal} and \ref{tab:scope}; every point contains $100$ independent trials with $K=100$, $N_{\mathrm{opt}}=1000$ and $N_{\mathrm{cert}}=10000$. Certification always uses the raw signed condition $S_d^{\mathrm L}>2$.}
\label{fig:certification-scope}
\end{figure*}

Figure~\ref{fig:certification-scope} shows that, for the three studied input-families, the unchanged protocol remains effective for maximally entangled reference inputs through $d=8$, while the observed affine-versus-independent contrast exposes the model capacity of four adjustable offsets. The dip in the maximally entangled certified rate at $d=3$ reflects $26$ of the $100$ qutrit trials ending in a lower optimisation basin, rather than any change in the direct protocol's detected-pair allocation.

Equation~\eqref{eq:affine-ramp} describes a phase misalignment between the incoming modes and the Fourier analyser, with the phase changing by a constant amount from one mode to the next. For each complete run, the source is assigned one randomly chosen (``unknown'') affine slope $\delta$, and the same value of $\delta$ is used for every detected pair throughout that run. At iteration $k$, CSPSA then uses the measured Bell value to update only the four Fourier measurement offsets $\bm\theta_k$. The affine-ramp results in Fig.~\ref{fig:certification-scope} show how often this measurement-side adjustment produces a certified violation through $d=8$. 

For isotropic white noise, Eq.~\eqref{eq:isotropic} makes the exact standard-Fourier value linear in $v$.  This family is sometimes referred to as Werner-type, for $d=2$. 
 Table~\ref{tab:white-noise} centres the finite-shot study on visibility $v_{\mathrm F}$. At the reference boundary the exact standard-Fourier value is $2$, and in this case, no run satisfies the   Hoeffding criterion. 
 A margin of $v-v_{\mathrm F}=0.02$ already yields certified rates of $43$--$62\%$ across $d=2$--$8$; at $0.05$ the rates are $71$--$100\%$. The visibility cases use independently generated runs, so small departures from a monotonic trend are ordinary Monte Carlo variation rather than changes observed within matched runs. 

\begin{table*}[t]
\centering
\setlength{\tabcolsep}{8pt}
\begin{tabular}{c@{\hspace{1.1em}}c@{\hspace{1.1em}}ccccc}
\toprule
$d$ & $v_{\mathrm F}$ & $v-v_{\mathrm F}=-0.02$ & $0$ & $+0.02$ & $+0.05$ & $+0.10$\\
\midrule
2 & 0.707107 & 0 & 0 & 62 & 100 & 100\\
3 & 0.696152 & 0 & 0 & 52 & 71 & 75\\
4 & 0.690550 & 0 & 0 & 50 & 82 & 85\\
5 & 0.687157 & 0 & 0 & 54 & 79 & 84\\
6 & 0.684884 & 0 & 0 & 58 & 82 & 77\\
7 & 0.683256 & 0 & 0 & 46 & 81 & 78\\
8 & 0.682033 & 0 & 0 & 43 & 87 & 84\\
\bottomrule
\end{tabular}
\caption{Hoeffding-certified runs out of $100$ for isotropic white noise, centred on the standard-Fourier reference visibility $v_{\mathrm F}=2/F_d(\bm\theta_\star)$. The independently seeded $v-v_{\mathrm F}=-0.05$ ensembles also give zero certified runs in every dimension. Each case uses $K=100$, $N_{\mathrm{opt}}=1000$ and $N_{\mathrm{cert}}=10000$.}
\label{tab:white-noise}
\end{table*}

\begin{figure*}[t]
\centering
\includegraphics[width=0.98\textwidth]{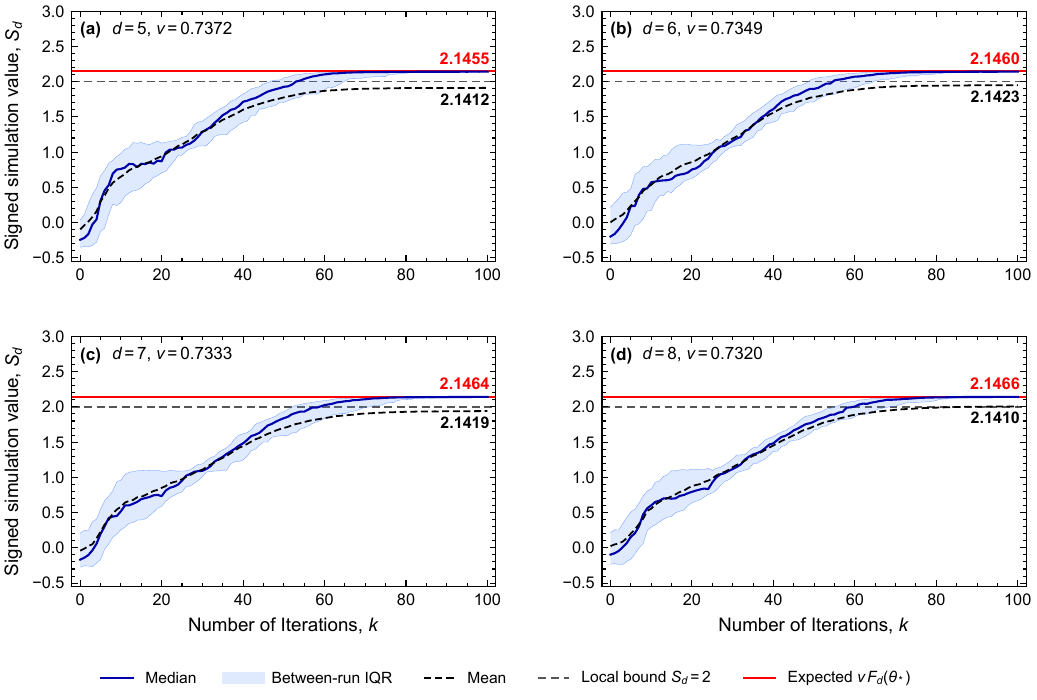}
\caption{Per-update phase-only CSPSA convergence for high-dimensional isotropic white-noise (Werner-type) inputs $\rho_v$. Panels (a)--(d) show $d=5,6,7,8$, respectively, at the common reference margin $v-v_{\mathrm F}=0.05$, where $v_{\mathrm F}=2/F_d(\bm\theta_\star)$ is the visibility for which the standard-Fourier Bell value equals the local bound. Each panel contains $100$ independently seeded $K=100$ trajectories with $N_{\mathrm{opt}}=1000$. The strong-blue curve is the across-run median of the exact signed central simulation value, the pale-blue band with fine boundaries encloses the first-to-third-quartile interval across runs, and the black dashed curve is the arithmetic mean over those $100$ runs. The grey dashed line is the local bound $S_d=2$; the red line and number give the exact standard-Fourier value $v\,F_d(\bm\theta_\star)$ for the mixed input, not a globally certified maximum over the four-offset Fourier measurement family. The central values are diagnostics \blue{evaluated at the phase setting reached at each iteration} and consume no additional pairs: the optimiser still uses only the two perturbed finite-shot Bell evaluations, and confidence certification uses four new terminal tables with $N_{\mathrm{cert}}=10000$.}
\label{fig:werner-convergence}
\end{figure*}

The smooth trajectories in Fig.~\ref{fig:werner-convergence} show how that mixed-state Bell margin is recovered rather than only reporting its terminal value. To compare dimensions at equal difficulty, each panel uses $v-v_{\mathrm F}=0.05$. Among successful runs, the median first exact crossing of the local bound occurs between updates $52$ and $59$ across $d=5$--$8$, and the final all-run median settles close to the exact standard-Fourier value $vF_d(\bm\theta_\star)$. The dashed mean remains lower because $13$--$21\%$ of the runs stay in the lower basin. This describes the all-run median when most runs occupy the upper branch. Some mixed-state runs are nevertheless unsuccessful.

\vspace{0.5\baselineskip}
\section{Closed form of the phase objective and Bell-operator interpretation}\label{sec:analytic}

For a maximally entangled input, the objective optimised by the phase-only protocol is
\begin{equation}
F_d(\bm\theta)=\langle\Psi_d|\mathcal B_d(\bm\theta)|\Psi_d\rangle.
\label{eq:tightnessF}
\end{equation}
Substitution of Eqs.~\eqref{eq:alicebasis} and \eqref{eq:bobbasis} gives
\begin{equation}
p(r,s|a,b)=\frac{1}{d^3}\left|\sum_{j=0}^{d-1}
e^{2\pi i j(s-r-\alpha_a-\beta_b)/d}\right|^2.
\end{equation}
Defining
\begin{equation}
Q_d(x)=\frac{1}{d^2}\frac{\sin^2(\pi x)}{\sin^2(\pi x/d)},
\label{eq:Qkernel}
\end{equation}
with its continuous value at removable singularities, and $\Sigma_{ab}=\alpha_a+\beta_b$, gives the closed-form Bell objective
\begin{widetext}
\begin{equation}
\begin{split}
F_d(\bm\theta)=\sum_{k=0}^{[d/2]-1}\left(1-\frac{2k}{d-1}\right)\{&Q_d(k+\Sigma_{00})+Q_d(k+1-\Sigma_{10})+Q_d(k+\Sigma_{11})+Q_d(k-\Sigma_{01})\\
&-Q_d(-k-1+\Sigma_{00})-Q_d(-k-\Sigma_{10})-Q_d(-k-1+\Sigma_{11})-Q_d(-k-1-\Sigma_{01})\}.
\end{split}
\label{eq:analyticFd}
\end{equation}
\end{widetext}
The sums obey $\Sigma_{00}+\Sigma_{11}=\Sigma_{01}+\Sigma_{10}$, leaving three independent combinations after fixing the common phase gauge.

At the standard Fourier point
\begin{equation}
\bm\theta_\star=(0,1/2,1/4,-1/4),
\label{eq:optimalphases}
\end{equation}
every positive term in Eq.~\eqref{eq:analyticFd} reduces to $Q_d(k+1/4)$ and every negative term to $Q_d(k+3/4)$. Its exact value is therefore
\begin{widetext}
\begin{equation}
F_d(\bm\theta_\star)=\frac{2}{d^2}\sum_{k=0}^{[d/2]-1}
\left(1-\frac{2k}{d-1}\right)
\left[
\csc^2\!\frac{\pi(k+1/4)}{d}
-\csc^2\!\frac{\pi(k+3/4)}{d}
\right].
\label{eq:Fdstandard}
\end{equation}
\end{widetext}
This expression evaluates the fixed-state objective exactly at $\bm\theta_\star$ for any $d$. Fixing the common gauge by setting $\alpha_0=0$, \blue{direct evaluation of the gradient and Hessian for $d=2$--$6$ gives} a zero gradient and three strictly negative Hessian eigenvalues at this point. Eight independently seeded searches over a complete period return the same $F_d$ value in each of those dimensions. The analytical calculation shows that the standard Fourier setting is a strict local maximum, while independent searches over a complete phase period consistently return the same Bell value, thereby providing reproducible numerical evidence. However, the result is specific to this measurement family and does not provide an upper bound for more general measurements.

Equation~\eqref{eq:Fdstandard} gives the maximum Bell value obtained within the four-offset Fourier measurement family when the input is fixed to the maximally entangled state. Ac\'{\i}n \emph{et al.}~\cite{Acin_QtmNonlocality} considered the broader Bell-operator optimisation in which the input state is also allowed to vary. For any fixed set of measurement offsets, the largest Bell value attainable over all input states is the largest eigenvalue of the corresponding Bell operator,
\begin{equation}
G_d(\bm\theta)=\lambda_{\max}\!\left[\mathcal B_d(\bm\theta)\right].
\end{equation}
This corresponds to the operator's highest-eigenvalue state, which is generally nonmaximally entangled for $d>2$ \cite{Acin_QtmNonlocality}. 

Table~\ref{tab:FG} compares the two values for \(d=2\)--\(6\), where both calculations were completed. In every case, four independent searches across the full phase range returned the value at the standard Fourier setting. 
Thus $F_d(\bm\theta_\star)$ is the standard maximally entangled-state CGLMP value, whereas $G_d(\bm\theta_\star)$ is aligned with the Ac\'{\i}n eigenvalue approach. The latter is larger for $d>2$ because the state is also selected, while that is not the objective maximised by our phase-only CSPSA.

\begin{table*}[t]
\centering
\setlength{\tabcolsep}{11pt}
\begin{tabular}{c@{\hspace{2em}}ccc}
\toprule
$d$ & $F_d(\bm\theta_\star)$ & $G_d(\bm\theta_\star)$ & Deterministic local range\\
\midrule
2 & 2.8284271247 & 2.8284271247 & $[-2,2]$\\
3 & 2.8729340512 & 2.9148542155 & $[-4,2]$\\
4 & 2.8962432185 & 2.9726982671 & $[-10/3,2]$\\
5 & 2.9105448081 & 3.0157104755 & $[-3,2]$\\
6 & 2.9202036064 & 3.0497004192 & $[-14/5,2]$\\
\bottomrule
\end{tabular}
\caption{\blue{Fixed-state and Bell-operator results for $d=2$--$6$.} $F_d$ is the exact signed expectation for the fixed maximally entangled state at the standard Fourier point. $G_d$ is the largest eigenvalue of the corresponding Bell operator and follows the Ac\'{\i}n-type state-optimised construction. Complete-period numerical searches reproduce the displayed candidates but are not used as global upper certificates.}
\label{tab:FG}
\end{table*}

The asymmetric local range makes the signed convention essential. For qutrits, enumeration of the $3^4$ deterministic response functions gives $-4\leq S_3\leq2$; a simulated score $S_3\simeq-3.761$ is therefore locally allowed even though its magnitude exceeds $3$. 

\section{High-dimensional physical-multinomial tomography benchmark and resource tradeoff}\label{sec:qst}

The direct protocol answers a decision problem of whether the returned settings certify $S_d>2$ without reconstructing the state. To quantify what is gained and what is traded away, we simulate a physical-multinomial, informationally complete QST comparator on the same affine-ramp inputs. To ensure that the comparison is task matched, both routes receive the same total detected-pair cap, infer settings for the same four-setting-pair CGLMP test, and finish with four fresh CGLMP tables evaluated by the same signed Hoeffding rule. 
\subsection{Physical acquisition and arbitrary-dimensional reconstruction}

The implemented scheme uses the $d^2-1$ traceless Hermitian generalised Gell--Mann observables $\{\lambda_i\}$, normalised by
\begin{equation}
\operatorname{Tr}(\lambda_i\lambda_j)=2\delta_{ij}.
\label{eq:gell-mann-normalisation}
\end{equation}
For every ordered local-observable pair $(i,j)$, write the spectral decompositions as $\lambda_i=\sum_r l^{(i)}_r\Pi^{(i)}_r$ and $\lambda_j=\sum_s l^{(j)}_s\Pi^{(j)}_s$. One complete physical $d\times d$ joint-outcome table is then simulated by a single multinomial draw,
\begin{align}
\bm n^{(ij)}&\sim\operatorname{Multinomial}
\left(N_{\mathrm{tomo}},\{p^{(ij)}_{rs}\}_{r,s=0}^{d-1}\right),\label{eq:qst-multinomial}\\
p^{(ij)}_{rs}&=\operatorname{Tr}\!\left[
\rho\left(\Pi^{(i)}_r\otimes\Pi^{(j)}_s\right)\right].\label{eq:qst-born}
\end{align}
This complete-table model preserves the covariance among outcomes from one setting. It differs from treating individual projectors as independent Bernoulli experiments.

The correlation moment is estimated directly from the table,
\begin{equation}
\widetilde T_{ij}=\sum_{r,s}l^{(i)}_r l^{(j)}_s
\frac{n^{(ij)}_{rs}}{N_{\mathrm{tomo}}}.
\label{eq:qst-correlation}
\end{equation}
Local moments are obtained by marginalising the same joint tables and averaging their repeated appearances over the other party's settings; no additional marginal acquisitions are charged. If $\widetilde a_i$, $\widetilde b_j$ and $\widetilde T_{ij}$ denote the resulting moments, arbitrary-dimensional bipartite linear inversion gives
\begin{equation}
\begin{split}
\rho_{\mathrm{LI}}={}&\frac{\mathbb I_{d^2}}{d^2}
+\frac{1}{2d}\sum_i\left(\widetilde a_i\lambda_i\otimes\mathbb I
+\widetilde b_i\mathbb I\otimes\lambda_i\right)\\
&+\frac14\sum_{i,j}\widetilde T_{ij}\lambda_i\otimes\lambda_j.
\end{split}
\label{eq:qst-li}
\end{equation}
\blue{The coefficients $\widetilde a_i$, $\widetilde b_j$ and $\widetilde T_{ij}$ are the generalised Stokes--Bloch components inferred from the complete joint tables, and $\rho_{\mathrm{LI}}$ is the corresponding direct linear reconstruction.}
Appendix~\ref{app:qst-derivation} gives the complete operator basis and the corresponding $n$-qudit expansion.

Since, finite count effects can make $\rho_{\mathrm{LI}}$ nonpositive, the reconstruction first Hermitises it and then performs the deterministic spectral projection
\begin{align}
\rho_{\mathrm H}&=\frac{\rho_{\mathrm{LI}}+\rho_{\mathrm{LI}}^\dagger}{2}
=\sum_q\eta_q|u_q\rangle\langle u_q|,\label{eq:qst-hermitise}\\
\rho_{\mathrm{PSD}}&=
\frac{\sum_q\max(\eta_q,0)|u_q\rangle\langle u_q|}
{\sum_q\max(\eta_q,0)}.\label{eq:qst-psd}
\end{align}
Thus every reported estimate is Hermitian, positive semidefinite and trace one. Equations~\eqref{eq:qst-hermitise} and \eqref{eq:qst-psd} define a physicality-enforcing projection, not maximum-likelihood reconstruction. \blue{For completeness, Appendix~\ref{app:qst-physical-fit} describes a dimension-generic triangular physical-state parametrisation and the associated likelihood-based alternatives~\cite{qtmtomo_qubits,qtmtomo_qudits}.}

\blue{The multinomial measurement record is fixed before reconstruction, so alternative classical estimators could be applied to the same count tables without further acquisition. We use the explicit positive-part renormalisation in Eq.~\eqref{eq:qst-psd} because it is deterministic and computationally inexpensive for the repeated high-dimensional reconstructions required here; related spectral-truncation and projected-linear approaches are discussed in Refs.~\cite{GranadePracticalAdaptiveTomography2017,SmolinFastMLE2012,GutaProjectedLeastSquares2020}. }

\subsection{Affine-phase inference and fresh certification}

For the affine state in Eq.~\eqref{eq:affine-ramp}, adjacent correlated coherences obey
\begin{equation}
\rho_{jj,(j+1)(j+1)}=\frac{1}{d}
\exp\!\left(-\frac{2\pi i\delta}{d}\right),
\qquad j=0,\ldots,d-2.
\label{eq:qst-affine-coherence}
\end{equation}
Their common argument therefore supplies a task-specific estimator of the unknown slope,
\begin{equation}
\blue{\widetilde{\delta}}=\operatorname{mod}_d\!\left[-\frac{d}{2\pi}
\arg\left(\frac{1}{d-1}\sum_{j=0}^{d-2}
\widehat\rho_{jj,(j+1)(j+1)}\right)\right].
\label{eq:qst-ramp-estimator}
\end{equation}
\blue{Because $\delta$ is defined modulo $d$, we report the circular slope error $e_\delta=\min\!\left\{\operatorname{mod}_d(\widetilde{\delta}-\delta),\operatorname{mod}_d(\delta-\widetilde{\delta})\right\}$, which is the shorter of the two distances between the reconstructed and true slopes on a circle of period $d$.}
The cyclic $j=d-1\rightarrow0$ coherence is excluded because its phase wraps differently under the chosen representative of $\delta$. The standard Fourier analyser is shifted by $(\blue{\widetilde{\delta}},\blue{\widetilde{\delta}},0,0)$ and is then tested with four \emph{new} multinomial tables. 
These certification counts are independent of the tomography data and use the same $N_{\mathrm{cert}}$ and the same criterion in Eq.~\eqref{eq:hoeffding} as CSPSA. This task-matched coherence estimator replaces any need to assert that a generic reconstructed pure target state can be reached from $|\Psi_d\rangle$ by local unitaries; for pure bipartite states, this is possible only when their Schmidt spectra agree.

\subsection{Matched setting and detected-pair budgets}

The direct resources are Eqs.~\eqref{eq:config-resource} and \eqref{eq:pair-resource}. \blue{For the QST route, the tomography-setting count and the total joint-setting count are}
\begin{equation}
M_d=(d^2-1)^2,\qquad
C_{\mathrm{set}}^{\mathrm{QST}}=M_d+4
\label{eq:qst-setting-budget}
\end{equation}
\blue{Here $M_d$ is the number of joint tomography settings in local dimension $d$: Alice and Bob each have $d^2-1$ Gell--Mann observables, giving $M_d=(d^2-1)^2$ ordered setting pairs. The symbol $C_{\mathrm{set}}^{\mathrm{QST}}$ denotes the total number of joint-setting configurations used in one QST attempt; it consists of the $M_d$ tomography settings and the four final CGLMP Bell-test settings. The superscript ``QST'' identifies the tomography route, while the subscript ``set'' indicates that this is a setting count rather than a detected-pair count.}

\blue{The symbol $B$ denotes the common total detected-pair cap for one attempt, and $N_{\mathrm{cert}}$ is the number of detected pairs reserved for each of the four final Bell-test settings.} The per-table allocation, acquired cost and unused integer remainder are
\begin{align}
N_{\mathrm{tomo}}(d)&=\left\lfloor
\frac{B-4N_{\mathrm{cert}}}{M_d}\right\rfloor,\label{eq:qst-floor}\\
B_{\mathrm{QST}}^{\mathrm{acq}}&=M_dN_{\mathrm{tomo}}+4N_{\mathrm{cert}},
\qquad R_d=B-B_{\mathrm{QST}}^{\mathrm{acq}}.\label{eq:qst-pair-budget}
\end{align}
\blue{Here $N_{\mathrm{tomo}}(d)$ is the integer number of detected pairs allocated to each of the $M_d$ tomography settings. The floor symbol ensures that this allocation is an integer and that the common cap is not exceeded. The quantity $B_{\mathrm{QST}}^{\mathrm{acq}}$ is the total number of pairs actually acquired by the QST route: $M_dN_{\mathrm{tomo}}$ pairs for tomography plus $4N_{\mathrm{cert}}$ pairs for the final Bell test. Finally, $R_d$ is the unused remainder, namely the difference between the allowed cap $B$ and the number actually acquired.}
For the fixed dimension-wide comparison in Table~\ref{tab:qst-acquisition}, $B=840000$ and $N_{\mathrm{cert}}=10000$. The table reports the exact integer allocations rather than treating the common cap as if it were always fully spent; Fig.~\ref{fig:qst-budget-benchmark} separately scans $B$ at $d=2,3$.

\begin{table*}[t]
\centering
\scriptsize
\setlength{\tabcolsep}{5.2pt}
\begin{tabular}{c@{\hspace{0.8em}}rrrrrrr}
\toprule
$d$ & $M_d$ & $N_{\mathrm{tomo}}$ & \shortstack{QST\\config.}
& \shortstack{acquired\\pairs} & $R_d$
& \shortstack{mean circular\\slope error} & \shortstack{median circular\\slope error}\\
\midrule
2 & 9    & 88888 & 13   & 839992 & 8    & $5.36\times10^{-4}$ & $3.86\times10^{-4}$\\
3 & 64   & 12500 & 68   & 840000 & 0    & $1.30\times10^{-3}$ & $1.04\times10^{-3}$\\
4 & 225  & 3555  & 229  & 839875 & 125  & $3.20\times10^{-3}$ & $2.80\times10^{-3}$\\
5 & 576  & 1388  & 580  & 839488 & 512  & $5.51\times10^{-3}$ & $4.55\times10^{-3}$\\
6 & 1225 & 653   & 1229 & 839925 & 75   & $8.79\times10^{-3}$ & $7.41\times10^{-3}$\\
7 & 2304 & 347   & 2308 & 839488 & 512  & $1.68\times10^{-2}$ & $1.38\times10^{-2}$\\
8 & 3969 & 201   & 3973 & 837769 & 2231 & $2.04\times10^{-2}$ & $1.93\times10^{-2}$\\
\bottomrule
\end{tabular}
\caption{Simulated physical-multinomial QST acquisition and affine-slope reconstruction under the common $840000$-pair cap. $M_d$ counts tomography setting pairs only; the configuration total includes four independent terminal Bell settings. Circular errors are in the dimensionless slope coordinate $\delta$ modulo $d$. The unused remainder is retained explicitly after the integer floor in Eq.~\eqref{eq:qst-floor}.}
\label{tab:qst-acquisition}
\end{table*}

To compare an imperfect decision procedure with a route that succeeds on every sampled input, we also report the point success-adjusted setting cost
\begin{equation}
C_{\mathrm{set/success}}=\frac{C_{\mathrm{set}}}{\widetilde p_{\mathrm{cert}}}.
\label{eq:success-adjusted}
\end{equation}
If $[p_{\mathrm L},p_{\mathrm U}]$ is the Wilson interval for the certified rate, the transformed cost interval is $[C_{\mathrm{set}}/p_{\mathrm U},C_{\mathrm{set}}/p_{\mathrm L}]$; its endpoints reverse because cost decreases with success probability.

\subsection{Benchmarking outcome}

Table~\ref{tab:resources} shows that the QST route certifies all $700$ paired affine inputs, with a Wilson $95\%$ interval of $[0.963,1]$ in every dimension, while the direct $K=100$ route inherits the basin-dependent rates from Table~\ref{tab:scope}. \blue{For reproducibility, each QST trial retains its simulated physical count tables and acquisition seed.} All $700$ trials pass table-normalisation, source-state, Hermiticity, positive-semidefinite and unit-trace checks; each uses every ordered Gell--Mann pair once and a terminal sample independent of the reconstruction. The structural result follows directly: $804$ direct configurations become fewer than the implemented QST count at $d=6$.

\begin{table*}[t]
\centering
\setlength{\tabcolsep}{6.5pt}
\begin{tabular}{c@{\hspace{1em}}cc@{\hspace{1.5em}}cc@{\hspace{1.5em}}cc}
\toprule
$d$ & \shortstack{CSPSA\\config./attempt} & \shortstack{QST\\config./attempt}
& \shortstack{CSPSA\\certified} & \shortstack{QST\\certified}
& \shortstack{CSPSA expected\\config./success} & \shortstack{QST expected\\config./success}\\
\midrule
2 & 804 & 13   & 100\% & 100\% & 804.0  & 13\\
3 & 804 & 68   & 70\%  & 100\% & 1148.6 & 68\\
4 & 804 & 229  & 56\%  & 100\% & 1435.7 & 229\\
5 & 804 & 580  & 40\%  & 100\% & 2010.0 & 580\\
6 & 804 & 1229 & 36\%  & 100\% & 2233.3 & 1229\\
7 & 804 & 2308 & 29\%  & 100\% & 2772.4 & 2308\\
8 & 804 & 3973 & 28\%  & 100\% & 2871.4 & 3973\\
\bottomrule
\end{tabular}
\caption{Matched affine-ramp comparison at $K=100$ under a common cap of $840000$ detected pairs per attempt. Every entry is based on $100$ paired input states. Expected configurations per certified success divide the per-attempt count by the observed certification rate. QST final Bell-test certification is performed using separately acquired counts and is not inferred from reconstruction fidelity.}
\label{tab:resources}
\end{table*}

\begin{figure*}[t]
\centering
\includegraphics[width=0.98\textwidth]{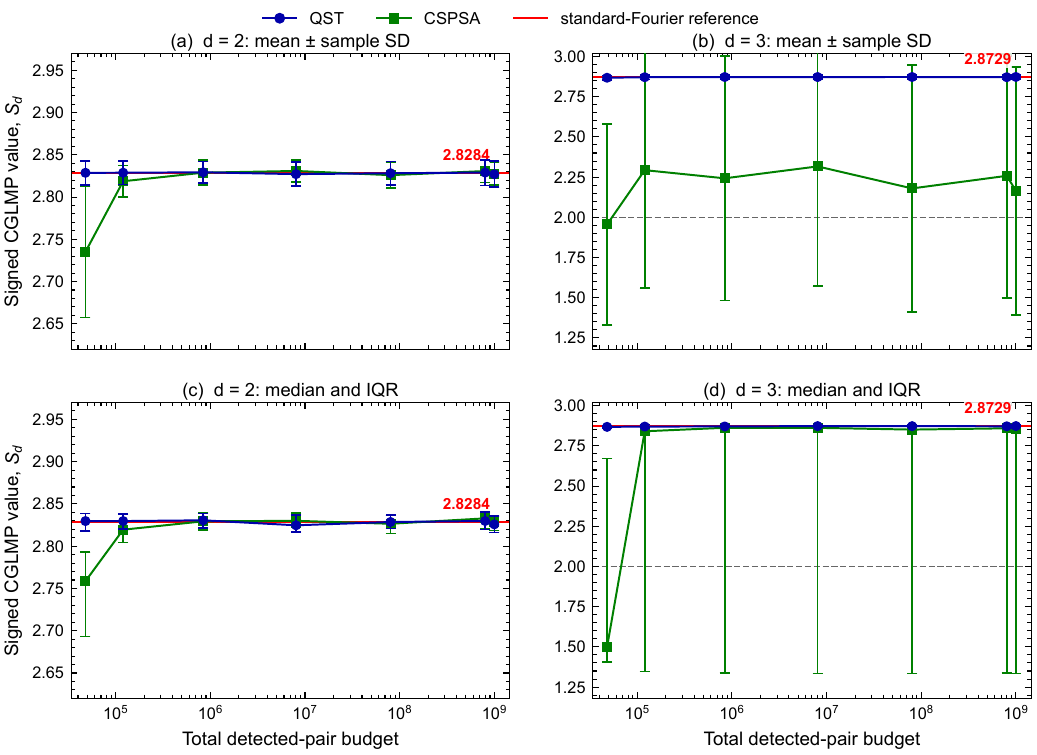}
\caption{Matched detected-pair-budget benchmark for affine-ramp inputs at $d=2$ and $3$. At each budget, the same $100$ frozen input identities are evaluated by statistically independent direct-CSPSA and QST routes. CSPSA uses $K=100$, the indicated $N_{\mathrm{opt}}$, and a fresh $N_{\mathrm{cert}}=10000$ acquisition, giving the cap $B=8KN_{\mathrm{opt}}+4N_{\mathrm{cert}}$. QST receives the same cap, allocates $\lfloor(B-4N_{\mathrm{cert}})/(d^2-1)^2\rfloor$ pairs to each complete Gell--Mann joint table, and finishes with four fresh terminal tables sampled from the true source state at settings inferred solely from the reconstruction; the integer remainder is unspent. Upper panels show the mean terminal signed $S_d$ with one between-run sample standard deviation, and lower panels show the median with the interquartile range. The red line is the exact standard-Fourier fixed-state reference and the grey line is the local bound. The horizontal coordinate is the declared common cap; lines guide the eye, and each budget uses independent acquisition streams for the same input identities. All Bell values are fresh finite-shot terminal samples; neither an absolute value nor a reconstructed-state surrogate is used.}
\label{fig:qst-budget-benchmark}
\end{figure*}

Figure~\ref{fig:qst-budget-benchmark} examines detected-pair dependence rather than only the main $840000$-pair point. \blue{At each of seven caps from $48000$ to $10^9$ pairs, the same $100$ affine-ramp input identities are used for both routes. Here ``frozen'' means only that this input ensemble was generated once and then held fixed across the budgets; the direct and tomography measurement records are sampled independently for every evaluation.} For $d=2$, QST has a resolved mean-score advantage at the two smallest caps: the paired QST-minus-CSPSA differences are $0.09351$ with bootstrap $95\%$ interval $[0.07863,0.10922]$ at $48000$ pairs, and $0.01010$ with interval $[0.00550,0.01465]$ at $120000$ pairs. From $840000$ pairs onwards every paired interval contains zero, and both routes certify all $100$ trials. For $d=3$, QST certifies all $100$ inputs at every budget, whereas direct CSPSA certifies $44$--$64$ because the fixed $K=100$ search repeatedly enters the lower basin. At $840000$ pairs this independently seeded sweep gives $59/100$ direct certifications, whereas the separately seeded fixed-budget ensemble in Table~\ref{tab:resources} gives $70/100$; their Wilson $95\%$ intervals, $[49.2,68.1]\%$ and $[60.4,78.1]\%$, overlap. At the same sweep point the paired mean difference is $0.62990$ with interval $[0.48907,0.78353]$. The direct median nevertheless lies near the standard-Fourier reference from $120000$ pairs onwards, which is why the figure reports both mean/standard deviation and median/IQR. This is a reliability advantage for the tailored affine-ramp QST comparator under the fixed direct protocol, not a statement that tomography is universally superior.

\begin{figure*}[t]
\centering
\includegraphics[width=0.94\textwidth]{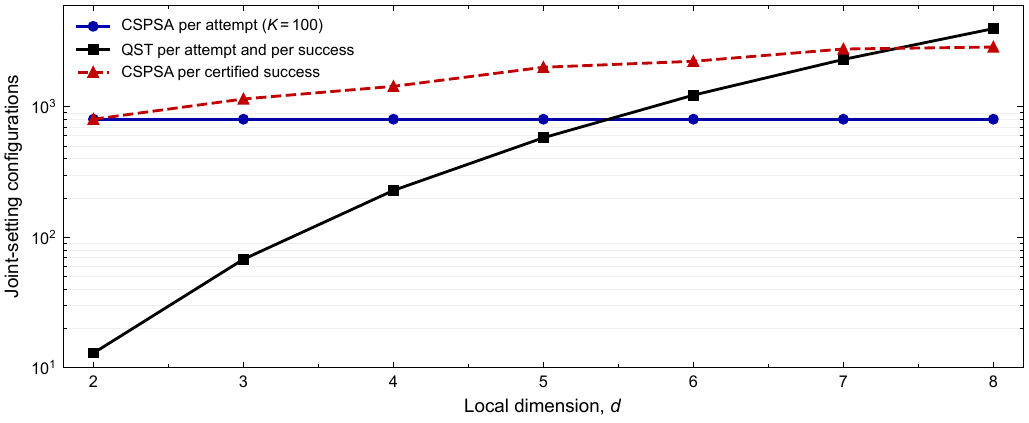}
\caption{Configuration tradeoff for the matched affine task at $K=100$, using Table~\ref{tab:resources}. CSPSA requires $804$ configurations per attempt, independent of $d$, while its success-adjusted curve divides this count by the observed certification rate. QST certifies every paired input, so its per-attempt and per-success curves coincide. The structural crossover occurs at $d=6$; the CSPSA success-adjusted point estimate crosses at $d=8$, without separation of the transformed $95\%$ intervals. Detected-pair costs are not plotted because no pair-efficiency crossover is observed through $d=8$.}
\label{fig:configuration-tradeoff}
\end{figure*}

At $K=100$, the point estimate of configurations per certified success first favours direct certification at $d=8$. The transformed $95\%$ intervals overlap, so this point-estimate crossover is not statistically separated. Figure~\ref{fig:configuration-tradeoff} shows this distinction from the structural $d=6$ crossover. With $K=50$, the affine certified rates are unchanged while the direct cost falls to $404$ configurations and $440000$ pairs per attempt. The reported $K=50$ sensitivity scan covers $d=5$--$8$: within that range, the point expected-configuration cost first favours CSPSA at $d=6$, the transformed intervals are disjoint from $d=7$, and the tailored QST comparator remains more pair-efficient. No $K=50$ affine comparison was run for $d=2$--$4$. Thus three statements must remain distinct: the structural configuration crossover is at $d=6$; the observed $K=100$ success-adjusted point crossover is at $d=8$; and, specifically for the matched $K=100$ comparison, no detected-pair crossover is observed through $d=8$.

\section{Discussion}\label{sec:discussion}

The central result is an adaptive Bell test that aligns its own high-dimensional Fourier phases and ends with a statistically valid signed certificate. The $S_+$ and $S_-$ acquisitions determine each update, and the independent four-setting acquisition at the returned phases is the final operational step. Optimisation and certification are therefore consecutive parts of one Bell-test procedure, with no density-matrix reconstruction.

The complete terminal ensembles reveal a multimodal landscape. Upper-branch runs reach the standard-Fourier benchmark across all dimensions, while a minority settle near much lower stationary values. This explains why the median can remain high when the mean falls, and why the pooled standard deviation does not shrink like a measurement error bar. The iteration scans show that the present failure probability is primarily a basin-selection problem: longer runs refine successful trajectories but do not materially change branch occupancy. Explicit multistart, restart or adaptive-gain variants are therefore the natural algorithmic developments, each with its own resource accounting and validation.

\blue{The affine-state comparison tests the reach of the four-offset Fourier measurement parametrisation. One phase slope can be cancelled by the measurement offsets, and for qubits this exhausts the diagonal relative-phase freedom. From qutrits upwards, generic independent phases contain additional degrees of freedom, and the observed certified rate declines accordingly. Table~\ref{tab:scope} quantifies this contrast: in one $100$-run ensemble per dimension, the affine-ramp certified counts for $d=2$--$8$ are $100$, $70$, $56$, $40$, $36$, $29$ and $28$, compared with $100$, $47$, $25$, $10$, $3$, $3$ and $0$ for generic independent phases. The protocol can be applied to general pure or mixed inputs; this comparison shows when the chosen four-offset analyser can remove unknown diagonal phase structure. Larger unknown basis transformations require a correspondingly larger orthogonality-preserving measurement family.}

Writing the objective in closed form allows us to see more clearly what the optimiser is able to find during its search. \blue{$F_d$ concerns a fixed maximally entangled input evaluated with the implemented four-offset Fourier measurements;} $G_d$ is an eigenvalue optimisation that changes the state selected by the Bell operator and is therefore the quantity aligned with the Ac\'{\i}n approach. Within its declared scope, the adaptive search recovers the positive standard-Fourier benchmark without knowing the phase alignment in advance.

The direct CSPSA protocol uses eight measurement configurations per iteration and bypasses state reconstruction. At the fixed $840000$-pair point the task-matched QST estimator succeeds more consistently and is more pair-efficient per certified success in the tested affine problem. The detected-pair sweep additionally shows no resolved qubit mean-score difference from that budget onwards and a qutrit reliability advantage for the tailored QST route because the direct search remains basin limited. In the reported higher-dimensional regimes, CSPSA uses fewer configurations per attempt from $d=6$, while its success-adjusted advantage depends on $K$ and the observed certification rate. The tomography comparison is itself fully dimension generic: it constructs all $d^2-1$ Gell--Mann observables, simulates one complete multinomial table for each ordered pair, reconstructs the bipartite state and enforces physicality before extracting the affine slope. Its role is  
to make the direct-versus-reconstruction benchmark physically specified, reproducible and operationally fair.  The phase-only CSPSA route is therefore especially attractive where changing a high-dimensional analyser is slow relative to accumulating additional photon pairs. 

Adaptive recovery of the available Bell margin may be useful for alignment in entanglement-based communication protocols \cite{E91_ekert,Durt_qutrits_theory_maxnoise}. The present evidence assumes known dimension, projective Fourier analysers and independently sampled pairs. Detector loss, basis-dependent efficiency, drift and imperfect outcome discrimination must be incorporated for a hardware-specific confidence statement \cite{DadaAndersson_OnBellViolations}. A nonviolating terminal test remains inconclusive about separability. 
\section{Conclusion}\label{sec:conclusion}

We have numerically demonstrated a self-aligning Bell-certification procedure for bipartite qudits by using the measured signed CGLMP value as the CSPSA objective. Two finite-shot multinomial Bell evaluations update all four phase offsets, and a fresh terminal acquisition supplies the confidence-certified result. With one unchanged phase-only algorithm, finite-shot certification is demonstrated through $d=8$. In one independently seeded ensemble of $100$ complete runs per dimension, the fresh Hoeffding test certified $100$, $74$, $84$, $80$, $80$, $78$ and $87$ maximally entangled inputs for $d=2$--$8$, respectively. Across $d=3$--$8$, the between-run standard deviation of $0.49$--$0.68$ is much larger than the mean terminal standard error of $0.011$--$0.013$; together with the iteration-budget comparisons, this identifies basin entry, rather than terminal shot uncertainty or insufficient run length, as the principal limit on reliability.

Affine Schmidt-phase ramps are exactly representable within the chosen four-offset Fourier measurements and can be aligned adaptively; generic independent phases quantify the boundary of that measurement parametrisation as dimension grows. In the corresponding $100$-run ensembles, the affine-ramp certified count decreases from $100$ at $d=2$ to $28$ at $d=8$, whereas the generic-independent-phase count decreases from $100$ to $0$. Under isotropic white noise, the exact standard-Fourier value scales linearly with visibility. Relative to the reference visibility $v_{\mathrm F}$, margins of $v-v_{\mathrm F}=0.02$ and $0.05$ give $43$--$62$ and $71$--$100$ certified runs per $100$, respectively, across $d=2$--$8$. The closed-form objective separates the fixed maximally entangled-state quantity $F_d$ from the Ac\'{\i}n-type eigenvalue quantity $G_d$ without claiming an unrestricted or globally certified maximum over the four-offset Fourier measurement family.

The operational cost is $8K+4$ configurations and $8KN_{\mathrm{opt}}+4N_{\mathrm{cert}}$ detected pairs. At the main operating point of $K=100$ iterations, $N_{\mathrm{opt}}=1000$ detected pairs per joint setting and $N_{\mathrm{cert}}=10000$ per fresh terminal joint setting, this is $804$ configurations and $840000$ detected pairs per attempt. The structural setting crossover occurs at $d=6$ and the success-adjusted point estimate at $d=8$, while no detected-pair crossover occurs through $d=8$. That is, although the success-adjusted configuration count favours CSPSA at \(d=8\), QST proves to be more efficient in terms of detected pairs per certified result throughout \(d=2\)--\(8\). The corrected $d=2,3$ budget sweep finds no resolved qubit mean-score difference from $840000$ pairs onwards and a qutrit reliability advantage for the tailored tomography route under the fixed $100$-iteration direct protocol. The matched comparator reconstructs all $700$ affine inputs with an arbitrary-dimensional physical-multinomial Gell--Mann scheme and certifies every corresponding returned setting on fresh simulated counts. The Bell test nevertheless answers the target question directly: it performs its own Fourier-phase alignment and ends in a fresh signed finite-shot certificate without first reconstructing a density matrix.

Future work will take this adaptive test towards experiment. Multistart or restart strategies can improve entry into the desired basin, while broader orthogonality-preserving analyser families can extend the search beyond four Fourier offsets. Incorporating detector loss, basis-dependent efficiency, drift and imperfect outcome discrimination will allow the finite-data certificate to be tested on real high-dimensional links \cite{DadaAndersson_OnBellViolations}. This is especially relevant to high-dimensional entanglement-based communication: in an E91-type QKD system, the Bell parameter directly tests the correlations shared across the link \cite{E91_ekert,Durt_qutrits_theory_maxnoise}. An analyser that aligns itself from the Bell data and then verifies the returned setting on a fresh sample could therefore become a useful part of practical Bell-certified high-dimensional links, and provide a starting point for future device-independent or semi-device-independent implementations.

\begin{acknowledgments}
This work was principally supported by a Royal Society Research Grant awarded to A.D. [Grant No. RG\textbackslash R1\textbackslash 251474]. J.M. acknowledges support from the UK Engineering and Physical Sciences Research Council (EPSRC) through the AQT Centre for Doctoral Training [EP/Y035089/1]. X.K.T. acknowledges sponsorship by the Ministry of Education, Singapore.
\end{acknowledgments}

\section*{Data availability}
The numerical data and code supporting the findings of this study are available from the corresponding author upon reasonable request.

\appendix

\section{Local range and projector checks}\label{app:checks}

For each $\bm\theta$, the numerical validation checks
\begin{align}
\sum_{r=0}^{d-1}M^A_{r|a}&=\mathbb I_d,&
\mathrm{Tr}(M^A_{r|a}M^A_{r'|a})&=\delta_{rr'},\\
\sum_{s=0}^{d-1}M^B_{s|b}&=\mathbb I_d,&
\mathrm{Tr}(M^B_{s|b}M^B_{s'|b})&=\delta_{ss'},
\end{align}
together with Hermiticity of $\mathcal B_d$ and normalisation of every $p(r,s|a,b)$. The fixed local range is obtained independently by enumerating deterministic response functions $(r_0,r_1,s_0,s_1)\in\{0,\ldots,d-1\}^4$ in Eq.~\eqref{eq:CGLMP Parameter}. The deterministic upper value is $2$ in every tested dimension. \blue{Direct enumeration gives the ranges shown in Table~\ref{tab:FG} for $d=2$--$6$, including the asymmetric qutrit interval $[-4,2]$.}

\begin{algorithm*}[t!]
\small
\DontPrintSemicolon
\caption{Matched simulated physical-multinomial QST affine-ramp comparator}
\label{alg:qst-comparator}
\KwIn{$d$, pair cap $B$, $N_{\mathrm{cert}}$, confidence level $1-\alpha$}
\KwOut{Physical estimate $\rho_{\mathrm{PSD}}$, slope $\blue{\widetilde{\delta}}$, fresh signed terminal certificate}
Construct the $d^2-1$ observables in Eqs.~\eqref{eq:gell-mann-s}--\eqref{eq:gell-mann-d}\;
Set $M_d=(d^2-1)^2$ and allocate $N_{\mathrm{tomo}}$ by Eq.~\eqref{eq:qst-floor}\;
\ForEach{ordered observable pair $(i,j)$}{
  Acquire one complete $d\times d$ multinomial table with $N_{\mathrm{tomo}}$ detected pairs\;
}
Estimate $\widetilde a_i,\widetilde b_j,\widetilde T_{ij}$ from Eqs.~\eqref{eq:qst-correlation}, \eqref{eq:qst-a} and \eqref{eq:qst-b}\;
Form $\rho_{\mathrm{LI}}$ and apply Eqs.~\eqref{eq:qst-hermitise} and \eqref{eq:qst-psd}\;
Estimate $\blue{\widetilde{\delta}}$ from Eq.~\eqref{eq:qst-ramp-estimator} and compensate the Fourier analyser\;
Acquire four \emph{new} Bell tables with $N_{\mathrm{cert}}$ detected pairs per setting\;
\Return{$\rho_{\mathrm{PSD}}$, $\blue{\widetilde{\delta}}$, the signed estimate, Hoeffding lower bound and decision $S_d^{\mathrm L}>2$}\;
\end{algorithm*}

\section{Generalised physical qudit tomography}\label{app:qst-derivation}

This appendix gives the arbitrary-dimensional operator construction underlying the benchmark in Sec.~\ref{sec:qst}. For $0\leq j<k\leq d-1$, the symmetric and antisymmetric off-diagonal generators are
\begin{align}
\lambda^{(\mathrm S)}_{jk}&=|j\rangle\langle k|+|k\rangle\langle j|,\label{eq:gell-mann-s}\\
\lambda^{(\mathrm A)}_{jk}&=-i|j\rangle\langle k|+i|k\rangle\langle j|.\label{eq:gell-mann-a}
\end{align}
The $d-1$ diagonal generators are
\begin{align}
\lambda^{(\mathrm D)}_{l}
&=\sqrt{\frac{2}{l(l+1)}}
\left(\sum_{j=0}^{l-1}|j\rangle\langle j|
-l|l\rangle\langle l|\right),\label{eq:gell-mann-d}\\[-2pt]
&\hspace{8em}l=1,\ldots,d-1.\notag
\end{align}
Together, Eqs.~\eqref{eq:gell-mann-s}--\eqref{eq:gell-mann-d} give
\begin{equation}
\frac{d(d-1)}2+\frac{d(d-1)}2+(d-1)=d^2-1
\end{equation}
traceless Hermitian operators satisfying Eq.~\eqref{eq:gell-mann-normalisation}.

An orthonormal operator representation convenient in arbitrary dimension is
\begin{equation}
Q_0=\frac{\mathbb I_d}{\sqrt d},\qquad
Q_i=\frac{\lambda_i}{\sqrt2},\quad i=1,\ldots,d^2-1,
\end{equation}
for which $\operatorname{Tr}(Q_\mu Q_\nu)=\delta_{\mu\nu}$. An arbitrary state of $n$ local qudits has the expansion
\begin{align}
\rho_{1\cdots n}
&=\sum_{\mu_1,\ldots,\mu_n=0}^{d^2-1}
t_{\mu_1\cdots\mu_n}
Q_{\mu_1}\otimes\cdots\otimes Q_{\mu_n},\label{eq:nqudit-expansion}\\
t_{\mu_1\cdots\mu_n}
&=\operatorname{Tr}\!\left[
\rho_{1\cdots n}
Q_{\mu_1}\otimes\cdots\otimes Q_{\mu_n}\right].\label{eq:nqudit-coeff}
\end{align}
For $n=2$, separating the identity, local and correlation components of Eqs.~\eqref{eq:nqudit-expansion} and \eqref{eq:nqudit-coeff} yields Eq.~\eqref{eq:qst-li}.

The actual marginal estimators reuse the physical joint tables:
\begin{align}
\widetilde a_i&=\frac{1}{d^2-1}\sum_{j=1}^{d^2-1}
\sum_{r,s}l^{(i)}_r\frac{n^{(ij)}_{rs}}{N_{\mathrm{tomo}}},\label{eq:qst-a}\\
\widetilde b_j&=\frac{1}{d^2-1}\sum_{i=1}^{d^2-1}
\sum_{r,s}l^{(j)}_s\frac{n^{(ij)}_{rs}}{N_{\mathrm{tomo}}}.\label{eq:qst-b}
\end{align}
Equations~\eqref{eq:qst-correlation}, \eqref{eq:qst-a} and \eqref{eq:qst-b} therefore require precisely $(d^2-1)^2$ complete joint-basis acquisitions. Every setting is counted once, and every outcome within a setting belongs to one common multinomial table.

\section{Positive parametrisation and a likelihood-based alternative}\label{app:qst-physical-fit}

\blue{A dimension-generic physical-state parametrisation of the type used in likelihood-based tomography~\cite{qtmtomo_qubits,qtmtomo_qudits} is defined as follows.} Let $D=d^n$ be the total Hilbert-space dimension and let $T(\bm t)$ be a complex lower-triangular $D\times D$ matrix,
\begin{equation}
T_{mn}=
\begin{cases}
t_m\in\mathbb R,&m=n,\\
x_{mn}+iy_{mn},&m>n,\\
0,&m<n.
\end{cases}
\label{eq:qst-T}
\end{equation}
The diagonal contributes $D$ real coefficients and the strict lower triangle contributes $D(D-1)$, giving
\begin{equation}
D+2\frac{D(D-1)}2=D^2
\label{eq:qst-T-count}
\end{equation}
real parameters. Consequently a single qudit uses $d^2$ coefficients and a bipartite $d\times d$ state uses $d^4$; trace normalisation removes one overall scale. The map
\begin{equation}
\rho(\bm t)=
\frac{T(\bm t)T^\dagger(\bm t)}
{\operatorname{Tr}[T(\bm t)T^\dagger(\bm t)]}
\label{eq:qst-cholesky}
\end{equation}
is Hermitian, positive semidefinite and trace one for every nonzero $\bm t$.

If this parametrisation is combined with the complete-table acquisition in Eq.~\eqref{eq:qst-multinomial}, the exact multinomial negative log-likelihood, up to constants independent of $\bm t$, is
\begin{equation}
-\log\mathcal L(\bm t)=
-\sum_{i,j,r,s}n^{(ij)}_{rs}
\log\!\left\{\operatorname{Tr}\!\left[
\rho(\bm t)\left(\Pi^{(i)}_r\otimes\Pi^{(j)}_s\right)\right]\right\}.
\label{eq:qst-nll}
\end{equation}
\blue{A James-type approximate-likelihood, or Pearson-weighted physical fit, provides an alternative way to use this positive parametrisation~\cite{qtmtomo_qubits}. Neither that approximation nor the exact likelihood in Eq.~\eqref{eq:qst-nll} is used for the benchmark reported here, which instead uses the complete multinomial acquisition, linear inversion and explicit spectral correction in Eqs.~\eqref{eq:qst-multinomial}--\eqref{eq:qst-psd}.}

\vspace{0.5\baselineskip}
\section{Operational protocols}\label{app:algorithm}

Algorithm~\ref{alg:phase-only-cspsa} gives the compact pseudocode used in the main text. For completeness, one operational trial is unpacked as follows.
\begin{enumerate}
\item Draw the initial real phase vector with $0\leq\theta_{0,i}<1$ for $i=1,\ldots,4$, and fix the gains in Table~\ref{tab:gains}.
\item At iteration $k$, draw $\bm\Delta_k\in\{-1,+1\}^4$ and form $\bm\theta_{k\pm}=\bm\theta_k\pm c_k\bm\Delta_k$.
\item Acquire four $d\times d$ multinomial count tables at each perturbed vector and calculate the signed values $S^+_{d,k}$ and $S^-_{d,k}$.
\item Form Eq.~\eqref{eq:gradient}, update with Eq.~\eqref{eq:update}, and optionally store $(S^+_{d,k}+S^-_{d,k})/2$ as the no-cost convergence proxy.
\item After $K$ updates, acquire four fresh terminal count tables at $\bm\theta_K$. Report the signed estimate from Eq.~\eqref{eq:physical-estimator}, its standard error from Eq.~\eqref{eq:bell-se}, the lower bound from Eq.~\eqref{eq:hoeffding}, and whether that bound exceeds $2$.
\end{enumerate}
The exact central objective is computed only for simulation diagnostics.

The matched tomography route is summarised in Algorithm~\ref{alg:qst-comparator}.

\bibliography{refs_v7}

\end{document}